\documentclass[epjST]{svjour}
\usepackage{graphics,epsfig}
\newcommand{\ci}[1]{\cite{#1}}

\newcommand{\ba}{\begin{eqnarray}}
\newcommand{\ea}{\end{eqnarray}}
\newcommand{\beqs}{\begin{eqnarray}}
\newcommand{\eeqs}{\end{eqnarray}}

\begin{document}
 \title{Some problems of determination of the spin structure of the scattering amplitude and experiments
 at NICA}
%\subtitle{Do you have a subtitle?\\ If so, write it here}
 \author{O.\,V. Selyugin\fnmsep\thanks{\email{selugin@theor.jinr.ru}  }
}
%\inst{3}
 \institute{JINR, Bogoliubov Laboratory of Theoretical Physics, 141980 Dubna, Moscow region, Russia }
 \abstract{The existing experimental data are examined under different assumptions about the structure of
 the scattering amplitude of the proton-proton and proton-antiproton elastic scattering at high energy
 to obtain the value of  $\rho(s,t)$, the ratio of the real to imaginary part  of the scattering amplitude in
 the Coulomb-hadron interference region. It is shown that the deviation of $\rho(s,t)$ obtained from
 the experimental data  of the proton-antiproton scattering at $3.8 < P_L <6.2\,$GeV/c from the
 dispersion analysis is concern in all examined assumptions.  }
 \maketitle
 \section{Introduction}

%	The dynamics of strong interactions  finds its most
%  complete representation in elastic scattering at small angles.
%  Only in this region of interactions we  can measure the basic properties that
%  define the hadron structure: the total cross section,
%  the slope of the diffraction peak and the parameter $\rho(s,t) $.
%  Their values
%  are connected, on the one hand, with the large-scale structure of hadrons and,
%  on the other hand, with the first principles which lead to the
%  theorems on the behavior of the scattering amplitudes
%  %at asymptotic   energies
%   \ci{mart,roy}.

   The measure of the $s$-dependence of the total cross sections $\sigma_{\rm tot}(s)$  and the value
   of $\rho(s,t)$, the ratio of the real to imaginary part of the  elastic scattering amplitude are very
   important as they are connected to each other by the integral dispersion relations \cite{disp-rel}.
%   \ba
%   \rho_{ \frac{pp}{p\bar{p}} }(E)  \sigma_{ \frac{pp}{p\bar{p}} }(E)
%   = && \frac{A_0}{\sqrt{E^2-m^2c^4}} \pm   \frac{A_1 E}{\sqrt{E^2-m^2c^4}}  \\ \nonumber
%   + && \frac{E^2}{\pi \sqrt{E^2-m^2c^4}}  \int_{m}^{\infty} dE^{/}  \sqrt{E^2-m^2c^4}
%   [\frac{ \sigma_{ \frac{pp}{p\bar{p}} }(E') }{E'(E'-E)} -
%     \frac{ \sigma_{ \frac{p\bar{p}}{pp} }(E'^{2}) }{E'^{2}(E'+E)}
%   \ea
    The validity of this relation can be check up at LHC energies. The deviation can point out the
    existence of the fundamental length at TeV order.  But for such a conclusion we should know
    with high accuracy the lower energy data as well.  At these energies we have many contributions
   to the hadron scattering  amplitudes coming from  the exchange of  different regions.  Now we
   cannot exactly calculate all  contributions and find their energy dependence. However, a great
   amount of  the experimental material allows us to make full  phenomenological analysis and obtain
   the size and form of the  different parts of the hadron scattering amplitude.  The difficulty
  is that we do not know the energy dependence of these amplitudes and  individual contributions
  of the asymptotic non-dying spin-flip  amplitudes. As was noted in \cite{wak}, the spin-dependent
  part of the interaction in $pp$ scattering is stronger than  expected and a good  fit to the data
  in the Regge model requires an enormous number of poles.

  In the general case, the determination of the total cross sections depends on the parameters of
 the elastic scattering amplitude: $\sigma_{\rm tot}$, $\rho(s,t)$, the Coulomb-nuclear interference
 phase $\varphi_{\rm cn}(s,t)$ and the elastic slope $B(s,t)$. For the definition of new effects
 at small angles and especially in the region of the diffraction minimum, one must know
 the effects of the Coulomb-nuclear interference with sufficiently high accuracy. The Coulomb-nuclear
 phase was calculated in the entire diffraction domain taking into account the form factors of the
 nucleons \cite{selmp1,selmp2,PRD-Sum}.

%
%Now we  do not know exactly, also from a theoretical
%  viewpoint,  the dependence of  different parts of the
%  scattering amplitude on $s$ and $t$. So, usually, we suppose
%   that the imaginary and real parts of the spin-nonflip
%  amplitude behave  exponentially  with the same slope, whereas the
%  imaginary and real parts of  spin-flip amplitudes, without the
%  kinematic factor $\sqrt{|t|}$, behave in the same manner with $t$ in the
%  examined domain of transfer momenta.
%  Moreover, we assume mostly  the energy independence of
%  the ratio of  spin-flip parts to  spin-non-flip parts of the
%  scattering amplitude.  All that is  our theoretical uncertainty.

 There was a significant difference between the experimental measurement of $\rho $, the ratio of
 the real part to the imaginary part of the scattering amplitude, between the UA4 and UA4/2
 collaborations at $\sqrt{s}=541 $\,GeV.
%As shown in Table 1, the resulting values for $\rho(0) $ appear inconsistent.
 A more careful analysis~\cite{SelyuginYF92,SelyuginPL94} shows that there is no contradiction
 between the measurements of UA4 and UA4/2. Now the present model gives for this energy
 $\rho(\sqrt{s}=541\,{\rm GeV}, t=0) = 0.163 $, so practically the same as in the previous
 phenomenological analysis.

 There are many experimental data on the elastic proton-proton and proton-antiproton scattering
 at non-high energies (at $10< p_L<100\,$(GeV/c). However, the extracted sizes of $\rho(s,t=0)$
 contradict each other in the different experiments and give the bad $\chi^2$ in the different
 models trying to describe the $s$-dependence of  $\rho(s,t=0)$ (see, for example, the results
 of the COMPETE Collaboration \cite{COMPETE1,COMPETE2}). A more careful analysis of these
 experimental data give in some cases an essentially different value of $\rho(s,t=0)$. For
 example, our analysis of the experimental data, which take into account the uncertainty of the
 total cross sections, gave the new sizes of $\rho(s,t=0)$, which differ from original experimental
 data by 25\,\% on average. For example, for $P_L = 19.23\,$GeV/c the experimental work gave
 $\rho(s,t=0)=-0.25 \pm 0.03$ and for $P_L = 38.01$\,GeV/c $\rho(s,t=0)=-0.17 \pm 0.03$. Our
 analysis gave for these values:  $\rho(s,t=0)=-0.32 \pm 0.08$ and $\rho(s,t=0)=-0.12 \pm 0.03$,
 respectively. This picture was confirmed by the independent analyses of the experimental data
 \cite{Kuznetzov} $52< p_L < 400$\,(GeV/c) of Fajardo \cite{Fajardo} and our. Both new analyses
 coincide with each other but differed from original experimental data.

%%==================Figs.2 =============================
% \label{sec:figures}
% \begin{figure}[h] \begin{center}
% \includegraphics[width=0.8\textwidth] {doc01b.eps}
% \end{center}
% \caption{The $\rho(s,t=0)$ - the ratio of the real to imaginary part of the elastic scattering
% amplitude of the proton-antiproton scattering at low energies. Line - the dispersion analysis
% \cite{Kroll}. Open cercles - the data with small errors \cite{Disser,Disser-data}. (Figure from
% \cite{Disser}). } \label{Fig_2}
% \end{figure}
%%%%%%%%%%%% FIG. 2 KONEC %%%%%%%%%%%%%%%%%%%%%%%%%%%%%%%%%%%%%%%%%%%%%%%%
%%%%%%%%%%%% %%%%%%%%%%%

  Of course, we have plenty of experimental data in the domain of small momentum transfer at
 low energies $3 < p_L < 12$\,GeV/c). With pity, most of these experimental data gave the
 large errors of the experimental data. At these energies, we have many contributions to the
 hadron-spin-flip amplitudes from different reggion exchanges. Now we cannot exactly calculate
 all contributions and find their energy dependence. But a great amount of the experimental
 material allows us to make complete phenomenological analysis and find the size and form of
 different parts of the hadron scattering amplitude. The difficulty is that we do not know
 the energy dependence of such amplitudes and individual contributions of the asymptotic
 non-dying spin-flip amplitudes.

 The data on the proton-proton elastic scattering at $3.7 < p_L < 6.2$\,(GeV/c) \cite{Disser,Disser-data} are most
 interesting. These experimental data have the high accuracies and give the extracted value
 of the $\rho(s)$ with high precision. 
 % On Fig.~1 these data are shown together with other
 %experimental data and the predictions of the dispersion analysis carried out by Kroll
 %\cite{Kroll}.
  The new data of $\rho(s)$ essentially differ from the theoretical analysis by the dispersion relations \cite{Kroll}.
 Hence, for the first time, a more careful analysis of the original experimental data is
 require, to take into account the different assumptions and corrections to the scattering
 amplitude.

 \section{Coulomb phase factor with hadron form-factor}

 The differential cross sections of the nucleon-nucleon elastic scattering can be written
 as the sum of the different spiral scattering amplitudes:
  % and spin correlation parameters are

 \begin{eqnarray}
  \frac{d\sigma}{dt} =
 \frac{2 \pi}{s^{2}}\left (|\phi_{1}|^{2} +|\phi_{2}|^{2} +|\phi_{3}|^{2}
  +|\phi_{4}|^{2}
  +4 | \phi_{5}|^{2}\right )\,.	\label{dsdt}
 \end{eqnarray}

 Every amplitude $\phi_{i}(s,t)$, including the electromagnetic and hadronic forces, can be
 expressed as
 \begin{eqnarray}
  \phi(s,t) =  F_{C} \exp{\left(i \alpha \varphi (s,t)\right)} + F_{N}(s,t)\,,
 \end{eqnarray}
 with
 \begin{eqnarray}
  \varphi(s,t) =  \varphi(t)_{C} - \varphi(s,t)_{CN}\,,
 \end{eqnarray}
 where $\varphi(t)_{C}$ appears in the second Born approximation of the pure Coulomb amplitude,
 and the term $\varphi_{CN}$ is defined by the Coulomb-hadron interference.

 If  the hadron amplitude is chosen in the standard Gaussian form $F_{N}(s,t) = h_{nf}(s) \
 \exp{(-B(s) q^{2}/2)}$, we can get a standard phase, which is used in most experimental works,
 \begin{eqnarray}
 \varphi(s,t) = \mp \left[\ln{(-B(s) t/2)} + \gamma\right],  \label{wyph}
 \end{eqnarray}
 where $-t=q^2$,  $B(s)/2$ is the slope of the nuclear amplitude, $\gamma$ is the Euler constant,
 and the upper (lower) sign corresponds to the scattering of particles with the same (opposite)
 charges.

 The influence of the electromagnetic form factor of scattered particles on
 % $\varphi_{C}$ and
 $\varphi_{CN}$ in the framework of the eikonal approach was examined by  Cahn \cite{can}.
 He derived for $t \rightarrow 0 $ the eikonal analogue \cite{wy} %(\ref{wy})
 and obtained the corrections
 \begin{eqnarray}
 \varphi_{CN} (s,t)&=&\mp [\gamma +\ln{ (B(s)|t| /2)} + \ln{ (1 + 8/(B(s)\Lambda ^2))} \nonumber\\
    & & + (4|t|/\Lambda ^2)\ \ln{ (4|t|/\Lambda^2)} + 2|t|/\Lambda^2]\,, \label{Chane-ph}
 \end{eqnarray}
 where $\Lambda$ is a constant entering into the dipole form factor. The calculations of the
 phase factor beyond the limit $t \rightarrow 0$ carried out in \cite{selmp1,selmp2,PRD-Sum}.

 The impact of the spin of scattered particles was analyzed in \cite{lap,bgl} by using the
 eikonal approach for the scattering amplitude. Using the helicity formalism for high
 energy hadron scattering in \cite{bgl} it was shown that at small angles, all the helicity
 amplitudes have the same $\varphi(s,t)$.

 \section{Impact of the CNI phase}

 Let us  determine the hadronic and electromagnetic spin-non-flip amplitudes as
 \begin{eqnarray}
  F^{h}_{\rm nf}(s,t) &=&  \left[\phi^h_{1}(s,t) + \phi^h_{3}(s,t)\right]/2; \ \ \ \
%      \\
 F^{c}_{\rm nf}(s,t)
 %   &=&
  = \left[\phi^{\rm em}_{1}(s,t) + \phi^{\rm em}_{3}(s,t)\right]/2; \label{non-flip}
  \end{eqnarray}
 and spin-flip amplitudes as
 \begin{eqnarray}
  F^{h}_{\rm sf}(s,t)  &=&  \phi^h_{5}(s,t) ; \ \ \ \
%      \\
 F^{c}_{\rm sf}(s,t)
 %   &=&
  =  \phi^{\rm em}_{5}(s,t).    \label{spin-flip}
  \end{eqnarray}
 Equation (\ref{non-flip},\ref{spin-flip}) was applied at high energies and at small
 momentum transfer, with the following usual assumptions for hadron spin-flip amplitudes:
%$\bullet$
 $\phi_{1}=\phi_{3}$, $\phi_{2}=\phi_{4} = 0 \ $;
%$\bullet$
 the slopes of the hadronic spin-flip and spin-non-flip amplitudes are equal.

 Let us make a new fit of the experimental data of the proton-antiproton scattering
 \cite{Disser,Disser-data} at low energies with different approximations of the
 Coulomb-hadron interference phase factor. First, we used a simple form of the phase,
 Eq. (\ref{wyph}). The obtained sizes of $\rho(s,t=0)$ are shown in Table 1.
 The results are distributed near  the sizes of $\rho(s)$ extracted during the
 experiments. In  two cases  they are only slightly above $\rho_{\rm exper.}$
 (at $P_L = 4.066, \ 5.94$\,GeV/c); in three cases they are more above
 $\rho_{\rm exper.}$ (at $P_L = 5.72, \ 6.23$\,GeV/c); and only in one case
 they lie low ((at $P_L = 3.7$\,GeV/c). If we take a more complicated phase,
 Eq.(\ref{Chane-ph}), the results of the fitting procedure will be practically
 the same (see Table 1). At last, if we use our phase, taking into account the
 two photon approximation and the dipole form factor, the new fitting procedure
 give the different sizes of $\rho(s)$ (the last coulomb of Table 1). Now the
 results lie above $\rho_{exper.}$ for all examined energies. Hence, the
 difference with the predictions of the dispersion analysis only increases.

   \phantom{.}
%\vspace{.5cm}
 \begin{table}[h] \centering
%\vspace{.5cm}
{ TABLE 1 \\
 Proton-antiproton scattering (the phase dependence)} \\

 \vspace{.1cm}
% \begin{table}
 \begin{tabular}{|c|c|c||c|c|c|} \hline
              &               &               &        &    &       \\
 \large{ $P_{L}$ }& \large{N} &\large{ $\rho_{\rm exp.}$ }&
 \large{ $\rho_{\varphi(\rm Born)}$ } & \large{ $\rho_{\varphi(\rm Cahn)}$ }& \large{$\rho_{\varphi(\rm our)}$ } \\
 (Gev/c) & & & & & \\                                               
              &               &               &        &    &       \\
 \hline
              &               &               &        &    &       \\
 3.702 & 34 & $+0.018 \pm 0.03$ &  $+0.0077 \pm 0.02$  &  $+0.0078 \pm 0.08$ & $+0.028 \pm 0.08$\\
 4.066 & 34 & $-0.015 \pm 0.03$ &  $+0.0377 \pm 0.02$  &  $+0.0378 \pm 0.08$ & $+0.0324 \pm 0.08$\\
 5.603 & 215&  $-0.047 \pm 0.03$ &  $+0.035 \pm 0.02$  &  $+0.036 \pm 0.08$ & $-0.0017 \pm 0.08$\\
 5.724 & 115&  $-0.051 \pm 0.03$ &  $+0.0139 \pm 0.02$  &  $+0.014 \pm 0.08$ & $-0.0088 \pm 0.08$\\
 5.941 & 140&  $-0.063 \pm 0.03$ &  $-0.0003 \pm 0.02$  &  $-0.004 \pm 0.08$ & $-0.0055 \pm 0.08$\\
 6.234 & 34 &  $-0.06 \pm 0.03$ &  $+0.0162 \pm 0.02$  &  $+0.0162 \pm 0.08$ & $-0.0216 \pm 0.08$ \\
              &               &               &        &    &       \\
 \hline
 \end{tabular}
% \caption{The $\rho(s,t=0)$ - the ratio of the real to imaginary part of the
% elastic scattering amplitude of the proton-proton scattering at low
% energies.
% }
 \label{Table-1}
 \end{table}
%\vspace{.5cm}

 \section{Impact of the spin-flip contribution}

  Usually, one makes the
  assumptions that the imaginary and real parts of the spin-non-flip
  amplitude have an exponential behavior with the same slope, and the
  imaginary and real parts of the spin-flip amplitudes, without the
  kinematic factor $\sqrt{|t|}$ \cite{sum-L},
   are proportional to the corresponding parts of the non-flip amplitude.
 For example, in \ci{akth}
  the spin-flip amplitude was   chosen in  the form
 \ba
       F_{h}^{fl}=\sqrt{-t}/m_{p} \ h_{sf} \ F_{h}^{nf}. \label{sflip}
 \ea
 That is not so as regards the $t$ dependence shown in Ref. \ci{soff}, where
 $F^{fl}_{h}$ is multiplied %the exponential form by the special function
 dependent on $t$. Moreover, one mostly  takes  the energy independence of
 the ratio of the spin-flip parts to the spin-non-flip parts of the
 scattering amplitude. All this is our theoretical uncertainty
 \cite{cudpr-epja,M-Pred}.

 According to the standard opinion, the hadron spin-flip amplitude is
 connected with the quark exchange between the scattering hadrons,
 and at large energy and small angles it can be neglected. Some models,
 which take into account the non-perturbative effects, lead to the
 non-dying hadron spin-flip amplitude \cite{mog2}. Another complicated
 question is related to the difference in phases of the spin-non-flip
 and spin-flip amplitude.

 In \cite{Runco}, it was shown that the analysis of the low energy
 experimental data does not reveal the impact of the contribution of
 the spin-flip amplitude on the extracted value of $\rho(s,t)$. Out
 opinion is that this result must be checked up, as at low energies
 the size of the spin-flip amplitude determined by the second region
 contributions has to be sufficiently large. In this work, we examine
 the simple model of the spin-flip amplitude and try to find its
 impact on the determination of $\rho(s,t)$ from the low energy data
 of proton-proton scattering. Hence, we take the spin-non-flip amplitude
 in the simplest exponential form
 \ba
       F_{h}^{nl}= h_{nf} \ [1+\rho(s,t=0)] \  e^{B t/2}.
 \ea
 and  the spin-flip amplitude in the form of eq. (\ref{sflip})
 The differential cross section in this case will be
 \begin{eqnarray}
 \frac{d\sigma}{dt} = 2 \pi \ [|\phi_{nf}|^{2} \ +  \ 2 |\phi_{sf}|^{2}]\,,
 \label{dsdt-sf}
 \end{eqnarray}
 where the amplitudes $\phi_{nf}$ and $\phi_{sf}$ include the corresponding
 electromagnetic parts and the Coulomb-hadron phase factors.

 The results of our new fits of the proton-antiproton experimental data
 at $ P_L = 3.7 .. 6.2 \ $GeV/c are presented in Table 2. The changes of
 $\sum_{i}\chi_{i}^2$ after including the contribution of the spin-flip
 amplitude is reflected as the coefficient
 \ba
 R_{\sum_{i}\chi_{i}^2} \ = \ \frac{\sum_{i}\chi^2_{i \ without  \ sf.}
 - \sum_{i}\chi^2_{i \ with  \ sf.} }{\sum_{i}\chi^2_{i \ without \ sf.}}.
  \label{Rchi}
 \ea

%\phantom{.}
%\vspace{.5cm}
 \begin{table}[h]  \centering
 { TABLE 2 \\
 Proton-antiproton elastic scattering (the spin dependence)} \\

 \vspace{.1cm}
 \begin{tabular}{|c|c|c||c|c|c|} \hline
              &               &               &          &    & \\
 \large{ $P_{L(GeV/c)}$ }  & \large{  N} &\large{ $\rho_{exp.}$ } &\large{$R_{\chi^2}$}
 &\large{  $\rho_{model}$} &\large{ $h_{spin-flip}$ } \\
              &               &               &          &    & \\ \hline
                  &               &               &          &    & \\
 3.702 & 34 & $+0.018 \pm 0.03$ &  $8 \%$  &  $+0.057 \pm 0.02$ & $ 49.8 \pm 1.4$\\
 4.066 & 34 & $-0.015 \pm 0.03$ &  $25 \%$  &  $+0.052 \pm 0.009$ & $ 48.9 \pm 0.7$\\
 5.603 & 215&  $-0.047 \pm 0.03$ &  $3.5 \%$  &  $+0.014 \pm 0.005$ & $ 35.6 \pm 4.$\\
 5.724 & 115&  $-0.051 \pm 0.03$ &  $6.5 \%$  &  $+0.023 \pm 0.004$ & $ 38.2 \pm 4.5$\\
 5.941 & 140&  $-0.063 \pm 0.03$ &  $4.5 \%$  &  $+0.007 \pm 0.003$ & $ 43.2 \pm 0.4$\\
 6.234 & 34 &  $-0.06 \pm 0.03$ &  $1. \%$  &  $ -0.016 \pm 0.001$ & $ 3.5 \pm 0.3$ \\
              &               &               &          &  &   \\
 \hline
 \end{tabular}
 \end{table}
%\vspace{.5cm}

 We again obtain the sizes of  $\rho$ near zero and most part positive.
 So the results do not come to the sizes are predicted by the dispersion
 analysis of \cite{Kroll}. Except the last energy the contribution of the
 spin-flip amplitude is filing and impact on the sizes of $\rho$. Most
 remarkable is that the obtained size of the constant of the spin-flip
 amplitude coincides for practically all examined energies, except the last.
 These constants have a lager size and sufficiently small errors. It is
 shown that in the careful analysis of the size of $\rho$ we need to take
 into account the contribution of the spin-flip amplitude to the differential
 cross section at small angles, at least at low energies. Of course, a more
 accurate analysis with the examined different $t$ dependence of the spin-flip
 amplitudes is needed.

 \section{Conclusion}

 The future new data from LHC experiments will give the possibility to carry out
 the new analysis of the  dispersion relation which can open a new effects, for
 example, a fundamental length order TeV. However, for such analysis one also
 needs the knowledge of the low energy data with  high accuracy. Now the existing
 forward-scattering data at $P_L=4 � 40$\,(GeV) of the size $\rho(s,t)$ contradict
 each other. The $\rho(s,t)$ - data of the proton-antiproton scattering at
 $P_L=3.7 � 6.2$\,(GeV) contradict the old dispersion analysis.

 The present analysis, which includes the contributions of the spin-flip amplitudes,
 also shows a large contradiction between the extracted value of $\rho(s,t)$ and
 the predictions from the analysis based on the dispersion relations. However, our
 opinion is that it an additional analysis is needed which will include additional
 corrections connected with the possible oscillation in the scattering amplitude and
 with the $t$-dependence of the spin-flip scattering amplitude. We hope that the
 forward experiments at NICA can give valuable information for the improvement of
 our theoretical understanding of the strong hadrons interaction. This is especially
 true for the experiment at NICA with polarized beams.

 \vspace{0.5cm}

 {\small The author would like to thank  J.R. Cudell for helpful discussions,
 gratefully acknowledges  the financial support from FRNS and  and would like to
 thank the  University of Li\`{e}ge where part of this work was done.   }

 \end{document}